\documentclass[12pt]{article}
\usepackage{amsfonts}
\usepackage{amssymb}
\usepackage{graphics,psboxit,amsmath}
\usepackage{subfigure}
\usepackage{graphicx}
\usepackage{verbatim}
\usepackage{epic}

\def\hybrid{\topmargin 0pt \oddsidemargin 0pt 
        \headheight 0pt \headsep 0pt
        \textwidth 16,5cm 
        \textheight 23cm 
        \marginparwidth .875in
        \parskip 5pt plus 1pt \jot = 1.5ex}

\hybrid

\catcode`\@=11

\def\marginnote#1{}
\newcount\hour
\newcount\minute
\newtoks\amorpm
\hour=\time\divide\hour by60
\minute=\time{\multiply\hour by60 \global\advance\minute by-\hour}
\edef\standardtime{{\ifnum\hour<12 \global\amorpm={am}%
        \else\global\amorpm={pm}\advance\hour by-12 \fi
        \ifnum\hour=0 \hour=12 \fi
        \number\hour:\ifnum\minute<10 0\fi\number\minute\the\amorpm}}
\edef\militarytime{\number\hour:\ifnum\minute<10 0\fi\number\minute}

\def\draftlabel#1{{\@bsphack\if@filesw {\let\thepage\relax
   \xdef\@gtempa{\write\@auxout{\string
      \newlabel{#1}{{\@currentlabel}{\thepage}}}}}\@gtempa
   \if@nobreak \ifvmode\nobreak\fi\fi\fi\@esphack}
        \gdef\@eqnlabel{#1}}
\def\@eqnlabel{}
\def\@vacuum{}
\def\draftmarginnote#1{\marginpar{\raggedright\scriptsize\tt#1}}

\def\draft{\oddsidemargin -.5truein
        \def\@oddfoot{\sl preliminary draft \hfil
        \rm\thepage\hfil\sl\today\quad\militarytime}
        \let\@evenfoot\@oddfoot \overfullrule 3pt
        \let\label=\draftlabel
        \let\marginnote=\draftmarginnote
   \def\@eqnnum{(\theequation)\rlap{\kern\marginparsep\tt\@eqnlabel}%
\global\let\@eqnlabel\@vacuum} }

\def\draft2{
        \def\@oddfoot{\sl preliminary draft \hfil
        \rm\thepage\hfil\sl\today\quad\militarytime}
        \let\@evenfoot\@oddfoot \overfullrule 3pt
        \let\label=\draftlabel
        \let\marginnote=\draftmarginnote
   \def\@eqnnum{(\theequation)\rlap{\kern\marginparsep\tt\@eqnlabel}%
\global\let\@eqnlabel\@vacuum} }

\def\preprint{\twocolumn\sloppy\flushbottom\parindent 2em
        \leftmargini 2em\leftmarginv .5em\leftmarginvi .5em
        \oddsidemargin -.5in \evensidemargin -.5in
        \columnsep .4in \footheight 0pt
        \textwidth 10.in \topmargin -.4in
        \headheight 12pt \topskip .4in
        \textheight 6.9in \footskip 0pt
        \def\@oddhead{\thepage\hfil\addtocounter{page}{1}\thepage}
        \let\@evenhead\@oddhead \def\@oddfoot{} \def\@evenfoot{} }

\def\numberbysection{\@addtoreset{equation}{section}
        \def\theequation{\thesection.\arabic{equation}}}

\def\underline#1{\relax\ifmmode\@@underline#1\else
        $\@@underline{\hbox{#1}}$\relax\fi}

\def\titlepage{\@restonecolfalse\if@twocolumn\@restonecoltrue\onecolumn
     \else \newpage \fi \thispagestyle{empty}\c@page\z@
        \def\thefootnote{\fnsymbol{footnote}} }

\def\endtitlepage{\if@restonecol\twocolumn \else \newpage \fi
        \def\thefootnote{\arabic{footnote}}
        \setcounter{footnote}{0}} 

\catcode`@=12
\relax

\def\figcap{\section*{Figure Captions\markboth
        {FIGURECAPTIONS}{FIGURECAPTIONS}}\list
        {Figure \arabic{enumi}:\hfill}{\settowidth\labelwidth{Figure
999:}
        \leftmargin\labelwidth
        \advance\leftmargin\labelsep\usecounter{enumi}}}
 \relax
\def\tablecap{\section*{Table Captions\markboth
        {TABLECAPTIONS}{TABLECAPTIONS}}\list
        {Table \arabic{enumi}:\hfill}{\settowidth\labelwidth{Table
999:}
        \leftmargin\labelwidth
        \advance\leftmargin\labelsep\usecounter{enumi}}}
 \relax
\def\reflist{\section*{References\markboth
        {REFLIST}{REFLIST}}\list
        {[\arabic{enumi}]\hfill}{\settowidth\labelwidth{[999]}
        \leftmargin\labelwidth
        \advance\leftmargin\labelsep\usecounter{enumi}}}
 \relax
\makeatletter
\newcounter{pubctr}
\def\publist{\@ifnextchar[{\@publist}{\@@publist}}
\def\@publist[#1]{\list
        {[\arabic{pubctr}]\hfill}{\settowidth\labelwidth{[999]}
        \leftmargin\labelwidth
        \advance\leftmargin\labelsep
        \@nmbrlisttrue\def\@listctr{pubctr}
        \setcounter{pubctr}{#1}\addtocounter{pubctr}{-1}}}
\def\@@publist{\list
        {[\arabic{pubctr}]\hfill}{\settowidth\labelwidth{[999]}
        \leftmargin\labelwidth
        \advance\leftmargin\labelsep
        \@nmbrlisttrue\def\@listctr{pubctr}}}
 \relax
\makeatother

\def\ba{\begin{equation}}
\def\ea{\end{equation}}

\def\no{\noindent}

\def\IR{\relax{\rm I\kern-.18em R}}

\begin{document}

\renewcommand{\theequation}{\thesection.\arabic{equation}}
\csname @addtoreset\endcsname{equation}{section}

\newcommand{\eqn}[1]{(\ref{#1})}
\newcommand{\be}{\begin{eqnarray}}
\newcommand{\ee}{\end{eqnarray}}
\newcommand{\non}{\nonumber}
\begin{titlepage}
\strut\hfill
\begin{center}

\vskip -1 cm


\vskip 2 cm

{\Large \bf First exact Geon found  is a non-singular  monopole,
propagating as a primordial gravitational pp-wave.} 

{\bf Nikolaos A. Batakis}

\vskip 0.2in
 
Department of Physics, University of Ioannina, \\
45110 Ioannina,  Greece\\
{\footnotesize{\tt (nbatakis@uoi.gr)}}\\
{\small July 10, 2014}

\end{center}

\vskip .4in

\centerline{\bf Abstract}

\no
Geons are particle-like electrovacua. The concept is well-defined, but it still lacks a proper first example. 
Emerging as such is a self-confined exact 2-parameter pp-wave non-Dirac monopole ${\cal G}$ with  primordial 
$Q/r^2$ ($r\geq r_o$) field plus higher moments. ${\cal G}$ has  effective mass, independently-scaled NUT-like charge $\kappa|Q|=2r_o$ as diameter, and  spin. ${\cal G}$ {\em cannot} have actual {\sc em} charge $Q$ 
(by $\partial{\cal G}=0$),  Ricci-flat limits, nor spacetime or Dirac-string singularities, but Dirac's quantization condition holds. ${\cal G}/2$, as an upgraded `Kerr-Newman'  alternative or ${\cal S}_Q$ geon, carries actual charge $Q$ confined by topology on a round-$S^2[r_o]$  physical singularity on $\partial{\cal S}_Q\neq0$. ${\cal G}$ and ${\cal S}_Q$ offer exact analytic models in particle physics and cosmology, notably for primordial gravitational waves, inflation, and pre-galactic dynamics.

\vskip 1cm
\no
\newline
\no
\newline
\vskip 2cm\no
[file: Geons]
\vfill
\no

\end{titlepage}
\vfill
\eject

\no
\section {Introduction}
 
A century-old interest on `small particles'
made of self-confined spacetime 
was alerted by Schwarzschild's 1915 solution and
evolved all the way into the 50s, with Einstein's own 
among widespread efforts to uncover
non-singular particle-like vacua or electrovacua
 \cite{e}\cite{c}\cite{l}. Epitomized as
{\em Geon} by Wheeler \cite{wh}, the
concept still lacks a proper first example,
namely a sufficiently stable and self-confined  exact
non-singular solution of Einstein's gravity coupled to 
sourceless Maxwell fields. The closest we'll 
ever come to exact {\em pure-vacuum} geons, which
would actually require exotic topologies \cite{l},
might well be the Taub-NUT (albeit effectively massless)  vacuum \cite{t}. 
This remarkable  space, tediously
assembled as {\sc nut-t-nut} from
the Taub and NUT vacua, and several decades since,
remains the only known exact
non-singular (and with no boundaries) Ricci-flat  space
with ${\cal S}=S^3\times\IR$ topology\footnote
{Taub and NUT sectors within  {\sc nut-t-nut} are
joined at $C^\infty$ junctions
across `Misner bridges' of former null squashed-$S^3$ boundaries.
These are physical (not mathematical `black-hole') singularities, namely
they have everywhere-regular Riemann tensor
and finite volume elements, 
in spite of geodesic incompleteness.
} 
\cite{rs}\cite{gp}. So, conclusively, we actually
have only approximations to desirable 4D geon
electrovacua \cite{ab}.
Meanwhile, the notorious lack of exact
solutions (plus concern for stability) has refocused interest
back to singular models via topological geons \cite{so},
and toward the quantum-mechanical 
properties of geon black holes or 
Reissner-Nordstr\"om versions of Taub-NUT \cite{gp}. 
Here, a 2-parameter family of primordial
self-confined pp-wave non-Dirac
monopoles with $Q/r^2$ ($r\geq r_o$) field,
the ${\cal G}={\cal S_-}\vee{\cal S_+}$, 
is proposed as the first exact geon.
The `${\cal G}/2$' or ${\cal S}_Q$ geon has actual $Q$-charge 
confined  topologically over a round $S^2[r_o]$ on
the $\partial{\cal S}_Q\neq0$
physical singularity\footnote 
{As we'll see, $2r_o=\kappa Q$ (a NUT-like charge) 
can be even smaller than Planck length, but it cannot vanish.}.

\no
Geons already have substantial applications, as noted.
However, if allowed (as a concept) to be singular,
they would have to be excluded from their expectedly
most important and natural presence, namely at the Big Bang,
immediately after the first quantum fluctuation(s) of
the vacuum. There, with inflatons as a suspended exception,
the only physical entity which could have existed
is the {\em graviton}
as a primordial gravitational-wave particle.
Proposed as analytic model for the latter, 
our primordial (Big-Bang) ${\cal G}_{\rm bb}$
pp-wave geon  will
be outlined as last example in the last section. So
we begin with the  Hilbert-Einstein Lagrangian for 
gravity, coupled by $\kappa$ to sourceless-Maxwell content in
\be
{\cal L}=\frac{1}{\kappa^2}\varepsilon_\alpha^{\;\beta}\,{\cal R}^\alpha_{\;\beta}-F\wedge\ast F\,,
\label{l}
\ee
to uncover ${\cal G}$ as a  particle-like
manifold with
electromagnetic ({\sc em}) content $F$. Symmetries make
${\cal G}$ a Bianchi-type IX (left-$SU(2)$ invariant), 
with an extra Killing vector $\partial_\psi$ for axial rotations $\psi\in[0,4\pi)$
as the only survivor of right-$SU(2)$ invariance. 
A non-singular ${\cal G}$ cannot carry {\em actual} mass $m_G$.
It can neither
have actual {\sc em} charge $Q$, 
by $\partial{\cal G}=0$ 
and $dF=d\ast F=0$ from (\ref{l}). 
Such aspects 
can here emerge only {\em a posteriori}, effectively or otherwise,
if at all.


\no
\section {A  preview of the geometry and content of ${\cal G}$}

\begin{figure}
\begin{center}
\includegraphics[width=16cm]{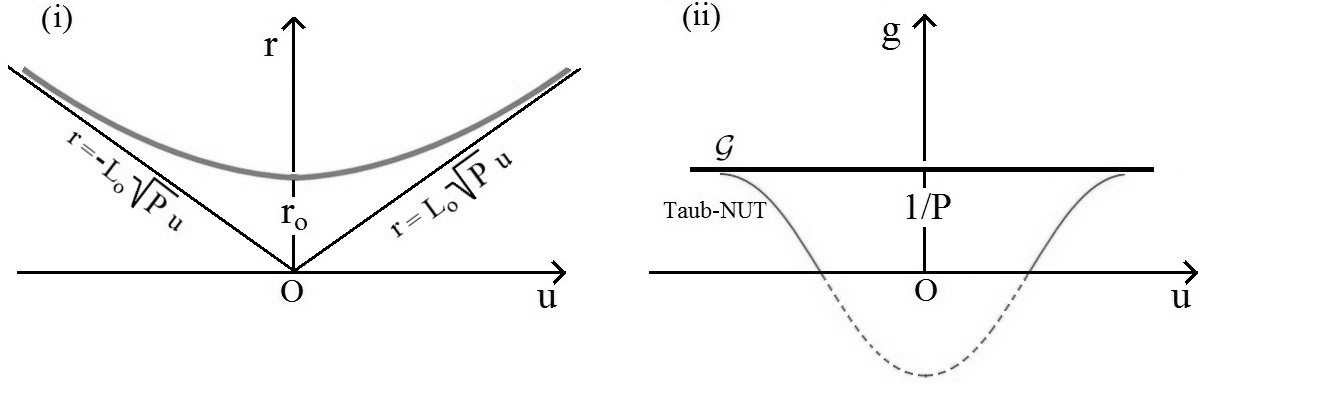}
\caption{${\cal G}$ vs Taub-NUT. 
(i) Both have the same $r=r(u)$ function on
$\pm{\rm L_o}\sqrt{P}u$ asymptotes
and $r_o$ minimum at $u=0$,  Planck-scaled
in ${\cal G}$ as a
$2r_o=\kappa|Q|$ NUT-like charge.
(ii) ${\cal G}$ has constant $g(u)=1/P$, thus
timelike $L_3$ $\forall u$,
hence no Taub sector (dashed) or Misner bridges.}
\label{Fig}\;
\end{center} 
\end{figure}

The  line element of ${\cal G}$ can be set
as a Taub-NUT type  in
terms of  left-$SU(2)$ invariant
1-forms $\ell^i$ (from $\theta, \phi, \psi$ angles on $S^3$),
scaled by ${\rm L_o}$, with $g=g(u)$, $r=r(u)$ functions of $u\in\IR$ in
\be
ds^2&=&-{\rm L_o^2}\left(g\ell^3+2du\right)\ell^3
+r^2d\Omega^2\,,\;\;\;\;
d\Omega^2\!\!:\,=(\ell^1)^2+(\ell^2)^2=d\theta^2+\sin^2\theta \;d\phi^2,\;\;\;
\label{tm}
\ee
\be
\ell^1=\cos\psi d\theta+\sin\psi\sin\theta\,d\phi,\;\;\;
\ell^2=-\sin\psi d\theta+\cos\psi\sin\theta\,d\phi, \;\;\;
\ell^3=d\psi+\cos\theta\,d\phi\,.
\label{l123}
\ee
The $L_i$, as duals of $\ell^i$ in
$<\!\!\ell^i|L_j\!\!>=\delta^i_j$, obey 
$<\!\!\ell^i|[L_j,L_k]\!\!>=\epsilon^i_{jk}$ via the equivalence of 
\be
d\ell^i\!=\!-\frac{1}{2}\epsilon^i_{jk}\ell^j\wedge\ell^k\;\;\;
\longleftrightarrow\;\;\;\;[L_j,L_k]=\epsilon^i_{jk}L_i\,,
\label{dual}
\ee
with, as we read off (\ref{tm}),
$(L_1)^2=(L_2)^2=r^2$, $(L_3)^2=-g{\rm L_o^{\,2}}$,
$\partial _u^{\,2}=0$,
$L_3\cdot\partial _u\!=\!-{\rm L_o^{\,2}}$, etc,
including the $\partial _u$ null vector.
Einstein's equations in orthonormal
Cartan frames  \cite{c}, to make  (\ref{tm}) locally manifest-Lorentz 
with $\eta_{\alpha\beta}={\rm diag}[-1,1,1,1]$
($\alpha=0,i$), will emerge as
\be
R_{\alpha\beta}=\kappa^2\,T_{\alpha\beta}^{\rm ({\sc em})},\;\;\;\;\;\;\;
T_{\alpha\beta}^{\rm ({\sc em})}=\frac{E^2+B^2}{2}\left[
\begin{array}{cccc}1&0&0&0\\0&1&0&0\\0&0&1&0\\0&0&0&-1\end{array}
\right],
\label{ee}
\ee
namely sourced by {\sc em}  content 
$T_{\alpha\beta}^{\rm ({\sc em})}$ in ${\cal G}$,
a manifest locally-Minkowski 
$M^4_{\!o}$ in those frames.
The scaleless $u\in\IR$ null coordinate
will be used as global time for now, to be shortly redefined 
as $t$ and later-on as $\rho$.
To preview the geometry and how ${\cal G}$
acquires particle-like {\em size}
from the $r=r_o$ minimum,
we need the $r=r(u)$, $g=g(u)$; they'll emerge from
(\ref{ee}) as
\be
r^2=r_o^2+{\rm L_o^2}Pu^2\,,\;\;\;
g=\frac{1}{P}\,,\;\;\;\;\;\;\;\;\;\;\;
\left[2r_o=\frac{\rm L_o}{\sqrt{P}}=\kappa|Q|\right],
\label{r}
\ee
in terms of the scaleless $P>0$ and $Q\neq0$
constant parameters. Thus,
${\cal G}$ and Taub-NUT
share {\em type} of metric, 
scale ${\rm L_o}$ and all isometries in (\ref{tm}), plus
invariance under 
\be
u\rightarrow -u\,,\;\;\;\;\;u\rightarrow u+u_o
\label{rt}
\ee
reflections and translations of $u$.
They also share the spacelike radius $r$ of $S^3$
as $r=r(u)$ function with $\pm{\rm L_o}\sqrt{P}u$ asymptotes
in (\ref{r}) and (\ref{as}), all depicted in diagram (i) of Fig.\ref{Fig}.
These strong similarities do not
inhibit stronger differences,
by which ${\cal G}$ cannot even  reduce to a Taub-NUT.
Actually, ${\cal G}$ is forbidden to reduce to {\em any}
Ricci-flat or singular limit, because the
$2r_o=\kappa|Q|$ NUT-like charge cannot vanish.
${\cal G}$ also carries the 
$\kappa^2=8\pi G_{\rm N}$ scale and   
the $Q$ (electric or magnetic) charge parameter,
with no counterpart in the also 2-parameter Taub-NUT. 
The latter's NUT charge as $2r_o={\rm L_o}/\sqrt{P}$
is fundamentally different
from the (formally identical) NUT-like charge
$2r_o={\rm L_o}/\sqrt{P}$ as a diameter in ${\cal G}$.
The point (and third difference) here is that  in ${\cal G}$
we {\em also} have $2r_o=\kappa|Q|$
(a geometric-mean of couplings,
if $Q^2\sim1/137$)
as  a second equivalent expression\footnote
{Taub-NUT carries no $\kappa$ scale or $Q$ charge, so
its NUT charge $2r_o={\rm L_o}/\sqrt{P}$ is unrelated to 
Planck scale etc, hence it is neither an {\em a priori} physical 
counterpart of $2r_o=\kappa|Q|$ in ${\cal G}$.
Thus, the {\em ad hoc} choice of $\kappa|Q|\sqrt{P}$
as ${\rm L_o}$ in one Taub-NUT
would append an {\sc em} aspect to the NUT charge
of that particular vacuum.
}.
The fourth difference is crucial:
$g(u)$  in ${\cal G}$ is by (\ref{r}) a $g=1/P>0$ constant,
approached only asymptotically by $g=g(u)$ in Taub-NUT,
as shown in diagram (ii) of Fig.\ref{Fig},
Thus, by $(L_3)^2=-(2r_o)^2$, $L_3$ 
is timelike everywhere
in ${\cal G}$. As a result, Taub sector and Misner bridges,
vital as they have been to keep Taub-NUT `standing', 
do not exist in ${\cal G}$, as if {\sc em} content had filled-in for `support'.
Accordingly, ${\cal G}$  consists of two  NUT-like pieces ${\cal S_\pm}$ in 
$C^\infty$ junction at $u=0$ 
(with ${\cal S_+}\leftrightarrow{\cal S_-}$ under $u\leftrightarrow-u$),
as a ${\cal G}={\cal S_-}\vee{\cal S_+}$,
sort of a {\sc nut-t-nut} with the Taub removed.
The ${\cal S_\pm}$ propagate as gravito-{\sc em}  solitons
along the null $\pm\partial_u$ wave vector, which
obeys the $D\partial_u=0$  pp-wave condition.
${\cal S_-}$ propagates backwards in time,
towards $r\rightarrow-{\rm L_o}\sqrt{P}u$ in (i) of Fig.\ref{Fig},
as an antisoliton. The inverse of $r=r(u)$ in (\ref{r})
is a double-valued  $u=u(r)$ function
\be
u=\pm\frac{1}{{\rm L_o}\sqrt{P}}\sqrt{r^2-r_o^2}\;\;\;\;\;
\stackrel{r>>r_o}{\longrightarrow}
\;\;\;\;\;\;\;\;\;u=\pm\frac{1}{{\rm L_o}\sqrt{P}}\;r
\;\;\rightarrow\;\;r={\rm L_o}\sqrt{P}\,|u|,
\label{as}
\ee
with a double-valued  limit at $r>>r_o$, so $r$ cannot cover 
${\cal G}$ globally. This gives rise to
the notion of a ${\cal G}/2=\!{\cal S}_Q$ geon as
a manifold covered globally by $r$, inevitably with
$\partial{\cal S}_Q\neq0$. This boundary at $r=r_o$
(the former junction at $u=0$) is
a squashed-$S^3$ physical singularity
with a round-$S^2[r_o]$
spacelike section of diameter $2r_o=\kappa|Q|$.
The  $r>>r_o$ limit in (\ref{as}) is a null-cone, depicted
in (i) of Fig.\ref{Fig} as
asymptotes, whose $r\geq 0$ range
in tangent space is clearly distinct from the $r\geq r_o$
range in ${\cal G}$ or ${\cal S}_Q$, wherein
the $r=r_o$ minimum is a perfectly regular point
(cf., next section). 
We can now trade manifest left-$SU(2)$
invariance in (\ref{tm}), using (\ref{l123}) and (\ref{as}),
for global $t$-time defined by
$\sqrt{P}t={\rm L_o}(\psi+Pu)$ in
\be
ds^2&=&-\left(dt+2r_o\cos\theta d\phi\right)^2
+\frac{r^2}{r^2-r_o^2}(dr)^2
+r^2\left(d\theta^2+\sin^2\theta d\phi^2\right),
\label{m0}
\\
dt&=&{\rm L_o}\left(\frac{1}{\sqrt{P}}d\psi+{\sqrt{P}}du\right)
=2r_o\left(d\psi+Pdu\right)\,.
\label{dt}
\ee
The {\sc em} content of gauge potentials and $F=dA$ fields
in ${\cal G}$ and
${\cal S}_Q$ is also non-singular everywhere (c.f., sections 3,4).
As first of two cases, an `electric' type $F^{\rm(e)}=dA^{\rm(e)}$ in
\be
A^{\rm(e)}=\frac{Q\sqrt{r^2-r^2_o}}{r^2}\,
\left(dt+2r_o\cos\theta\,d\phi\right)\;\longrightarrow\;\;
E^{\rm(e)}=\frac{Q}{r^2}+O(r^{-4})\,,\;\;B^{\rm(e)}\sim O(r^{-3}),
\label{eem}
\ee
in holonomic $(t,r,\theta,\phi)$ coordinates from (\ref{m0}), has
a dominant electric-monopole $E_C=Q/r^2$ field in $E^{\rm(e)}$ plus
electric and magnetic higher-order terms. Equally acceptable is the
\be
A^{\rm(m)}=-\frac{Q(r^2-2r_o^2)}{2r_or^2}\,
\left(dt+2r_o\cos\theta\,d\phi\right)\;\longrightarrow\;\;
E^{\rm(m)}\sim O(r^{-3})\,,\;\;B^{\rm(m)}=\frac{Q}{r^2}+O(r^{-4})\;\;
\label{mme}
\ee
`magnectic' type $F^{\rm(m)}=dA^{\rm(m)}$,
with a dominant magnetic-monopole $B_C=Q/r^2$
field in $B^{\rm(m)}$
plus higher-order terms, with
$F^{\rm(m)}=\ast F^{\rm(e)}$ by {\sc em} duality. 
We recall that actual charge $Q$
in ${\cal G}$ is forbidden by $dF=d\ast F=0$ and $\partial{\cal G}=0$,
so those $Q/r^2$  fields in ${\cal G}$ are {\em primordial}. 
${\cal S}_Q$, however, must carry actual 
surface-charge $\sigma_Q$ 
trapped by topology 
on a round-$S^2[r_o]$ in $\partial{\cal S}_Q\neq0$
(c.f., section 4). Particular choices of the $P,Q$ parameters 
can involve very different physical profiles
and  scales  in a  non-susy
hierarchy in the $r\geq r_o$ range of radii.
Depending on $|Q|$, we can have a
Planck-scale (or even smaller) $2r_o=\kappa|Q|\sim\kappa$
in a 4-scale hierarchy, up to a relatively enormous
$2r_o=\kappa|Q|\sim r_{\rm sm}$ minimum
in a 3-scale hierarchy, as
\be
r_o\sim\kappa|Q|\sim\kappa<<
r_{\rm sm}<<r_{\rm cl}<<r_\infty\,,\;\;\;\;\;\;\;\;\;\;
r_o\sim\kappa|Q|\sim
r_{\rm sm}<<r_{\rm cl}<<r_\infty\,,\;\;\;\;\;
\label{hi}
\ee
namely a Planck-length $r_o\sim\kappa$
(4-scale case),
vs a standard-model (below $10^{16}$ Gev) length
as $r_o\sim r_{\rm sm}$ (3-scale case);
common to both cases are
the $r\sim r_{\rm cl}$ scale
(a mean free path)
and $r\sim r_\infty$ (a Hubble radius) within a
Friedman model ${\cal F}$  filled with
${\cal G}$-geons.

\no
\section {The general  non-singular solution for ${\cal G}$}

We can  re-express 
(\ref{tm}) as $ds^2=\eta_{\alpha\beta}\theta^\alpha\theta^\beta$
with 1-forms  (plus duals) 
 in $<\!\!\theta^\alpha|\Theta_\beta\!\!>=\delta^\alpha_\beta$
as Cartan frames in non-singular geometry \cite{c},
which are chosen in terms of $du,\ell^i$ as
\be
\theta^0&=&{\rm L_o}\left(\sqrt{P}du+\frac{1}{\sqrt{P}}\ell^3\right),\;\;\;
\theta^1=r\ell^1,\;\;\;\theta^2=r\ell^2,\;\;\;
\theta^3={\rm L_o}\sqrt{P}du,
\label{cf}
\\
\Theta_0&=&\frac{\sqrt{P}}{\rm L_o}L_3,\;\;\;
\Theta_1=\frac{1}{r}L_1,\;\;\;\Theta_2=\frac{1}{r}L_2,\;\;\;
\Theta_3=\frac{1}{\rm L_o}\left(\frac{1}{\sqrt{P}}\partial_u-\sqrt{P}L_3\right),
\label{cF}
\ee
so they are also manifest left-$SU(2)$ invariant.
After the $D\theta^\alpha\!:\,=d\theta^\alpha+\Gamma^\alpha_{\;\beta}\wedge\theta^\beta=0$
definition for the covariant derivative,
we can easily verify the claimed $D\partial_u=0$ pp-wave condition,
while the (here non-holonomic) Christoffel 
$\Gamma^\alpha_{\;\beta\gamma}\theta^\gamma
=\Gamma^\alpha_{\;\beta}$ 1-forms follow as
\be
\Gamma_{\alpha\beta}&=&-\Gamma_{\beta\alpha}:\;\;\;\;\;\;\;\;\;\;
\Gamma_{01}=\frac{\rm L_o}{2\sqrt{P}r^2}\,\theta^2,\;\;\;\;
\Gamma_{02}=-\frac{\rm L_o}{2\sqrt{P}r^2}\,\theta^1,\;\;\;\;
\Gamma_{03}=0,\label{chris}
\\
\Gamma_{12}&=&-\frac{\sqrt{P}}{\rm L_o}
\left(1-\frac{\rm L_o}{2\sqrt{P}r^2}\right)\theta^0
+\frac{\sqrt{P}}{\rm L_o}\,\theta^3,\;\;\;\;
\Gamma_{23}=\frac{\dot r}{\rm L_o\sqrt{P}r}\,\theta^2,\;\;\;\;
\Gamma_{31}=-\frac{\dot r}{\rm L_o\sqrt{P}r}\,\theta^1,
\nonumber
\ee
with a dot for $d/du$. The curvature ${\cal R}^\alpha_{\;\beta}
=\frac{1}{2}R^\alpha_{\;\beta\gamma\delta}\theta^\gamma\wedge\theta^\delta
=d\Gamma^\alpha_{\;\beta}+
\Gamma^\alpha_{\;\gamma}\wedge\Gamma^\gamma_{\;\beta}$, which also 
supplies Ricci's $R_{\alpha\beta}=R^\gamma_{\;\alpha\gamma\beta}$,
gives Riemann's contractible components, all non-singular as
\be
R^0_{\,101}=R^0_{\,202}=-\frac{\rm L_o^2}{4Pr^4},\;
R^1_{\,212}=-\frac{1}{\rm L_o^2P}\left(\frac{\dot r}{r}\right)^{\!2}
+\frac{4Pr^2+3{\rm L_o^2}}{4Pr^4},\;
R^1_{\,313}=R^2_{\,323}=-\frac{1}{{\rm L_o^2}P}\frac{\ddot r}{r},
\label{riemann}
\ee
and Weyl's one independent 
component as $2PW_{0123}=-PW_{0312}=\dot r/r^3$,
vanishing as $O(r^{-3})$. By (\ref{l}),(\ref{ee}) etc,
$F=\frac{1}{2}F_{\alpha\beta}\,\theta^\alpha\!\wedge\theta^\beta$
can only have $F_{03}$ and $F_{12}$ components as
\be
F=-E\,\theta^0\wedge\theta^3+B\,\theta^1\wedge\theta^2\,,\;\;\;
\ast F=B\,\theta^0\wedge\theta^3+E\,\theta^1\wedge\theta^2\,.
\label{f}
\ee
With $\rho$-time  defined via
$r^2\dot\rho={\rm L_o^2}$ in $d\ast dA=0$, the
general solution is non-singular as
\be
A=\frac{Q\sqrt{P}}{\rm L_o}\sin(\rho+\rho_o)\;\theta^0\;,
\;\;\;\;\;\;\;\;\left[r^2d\rho={\rm L_o^2}du\right],
\label{a}
\ee
with $\rho_o$ a duality-rotation angle, supplying us with either of
$F^{\rm(e)}$ or $F^{\rm(m)}=\ast F^{\rm(e)}$ from
\be
A^{\rm (e)}=\frac{Q\sqrt{P}}{\rm L_o}\sin\rho\;\theta^0
\rightarrow\;\; F^{\rm (e)}=-\frac{Q}{r^2}\cos\rho\;\theta^0\wedge\theta^3
-\frac{Q}{r^2}\sin\rho\;\theta^1\wedge\theta^2,
\label{el}
\ee
\be
A^{\rm(m)}=-\frac{Q\sqrt{P}}{\rm L_o}\cos\rho\;\theta^0
\rightarrow\;\; F^{\rm(m)}=-\frac{Q}{r^2}\sin\rho\;\theta^0\wedge\theta^3
+\frac{Q}{r^2}\cos\rho\;\theta^1\wedge\theta^2.
\label{ma}
\ee
We always have $E^2+B^2=Q^2/r^4$, so we can  write-down and solve
(\ref{ee}) to establish (\ref{r}). For (\ref{eem}) and (\ref{mme}),
we first integrate $r^2\dot\rho={\rm L_o^2}$
to obtain $\rho=2\arctan{(2P u)}$, hence
\be
\sin\rho=\frac{2\tan{\rho/2}}{1+\tan^2\rho/2}=
\frac{2r_o\sqrt{r^2-r^2_o}}{r^2},
\;\;\;\;\;
\cos\rho=\frac{1-\tan^2\rho/2}{1+\tan^2\rho/2}=
\frac{r^2-2r_o^2}{r^2}.
\label{rho}
\ee
With these values used in (\ref{el}),(\ref{ma}), we find
the full result in  (\ref{eem}),(\ref{mme}) as
\be
A^{\rm(e)}=\frac{Q\sqrt{r^2-r^2_o}}{r^2}\,\theta^0\rightarrow\;\;
F^{\rm(e)}=-\left(\frac{Q}{r^2}-\frac{2Qr^2_o}{r^4}\right)\theta^0\wedge\theta^3
-\frac{2Qr_o\sqrt{r^2-r^2_o}}{r^4}\,\theta^1\wedge\theta^2,\;
\label{em}
\ee
\be
A^{\rm(m)}=-\frac{Q(r^2-2r_o^2)}{2r_or^2}\,\theta^0
\rightarrow\; F^{\rm(m)}=-\frac{2Qr_o\sqrt{r^2-r^2_o}}{r^4}\;\theta^0\wedge\theta^3
+\left(\frac{Q}{r^2}-\frac{2Qr^2_o}{r^4}\right)\theta^1\wedge\theta^2,\;\;\;
\label{me}.
\ee
We conclude  that all potentials and $E$,$B$ fields in ${\cal G}$ are
(i) non-singular $\forall \;r\geq r_o$, and
(ii) scaled by $\kappa$ via the $Q$ parameter {\em alone}; 
moreover, this {\sc em} content
(iii) can be directly read-off (\ref{em}) or (\ref{me}), and
(iv) indeed
includes a dominant Coulomb-like $E_C=Q/r^2$  electric
(or $B_C=Q/r^2$ magnetic) monopole field, 
a magnetic  dipole  moment ${\sf m}=2Qr_o$
in (\ref{em}) or electric ${\sf p}=2Qr_o$   in (\ref{me}),
plus quadrapole  moments.
To establish $E_C=Q/r^2$ as a
{\em primordial} field  in ${\cal G}$,
we apply the divergence theorem
(under $d\ast F=0$) in any $du=0$ 
hypersurface of simultaneity
with finite 3D volume ${\cal V}$.
\begin{figure}
\begin{center}
\includegraphics[width=12cm]{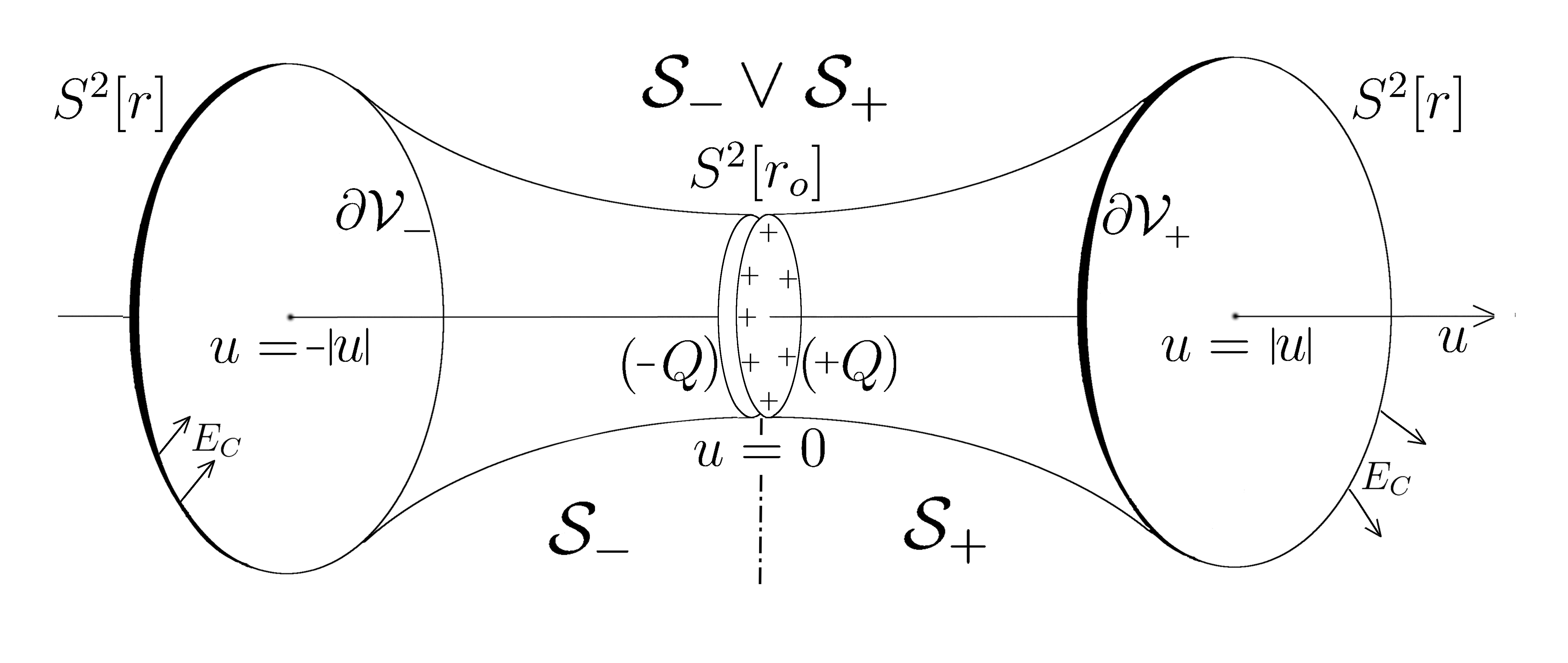}
\caption{Round $S^2[r]$  sections along the 
$r=\pm{\rm L_o}\sqrt{P}u$ null-cone as (not shown) asymptote.
(i) Disconnected $\partial{\cal V}_\pm$ boundary of $du=0$
hypersurface ${\cal V}$ in ${\cal G}$.
(ii) Via $S^2[r_o]$ and $S^2[r]$ on the right in ${\cal S_+}$
it is established that `${\cal G}/2$' $={\cal S}_Q$ carries actual $Q$ charge 
on the $S^2[r_o]$  of $\partial{\cal S}_Q$.}
\label{fig}\;
\end{center} 
\end{figure}
As usual, we can let $\partial{\cal V}$ 
surround the {\em origin}, which, instead of an ``$r=0$ point'',
is the $u=0$ locus, namely the small $S^2[r_o]$ sphere
shown in Fig.\ref{fig}.
By $\partial{\cal G}=0$
and the fact that such $u\!=$constant slices in ${\cal G}$
receive legitimate contributions from
{\em both} of  ${\cal S_\pm}$,
the $\partial{\cal V}$ of any $du=0$ volume ${\cal V}$
must be a disconnected  set.
This is actually shown in Fig.\ref{fig} as the
${\partial{\cal V_\pm}}$ pair of round-$S^2[r]$,
which approach asymptotically the (not shown) null cone
given as $r=\sqrt{P}{\rm L_o}|u|$ in (\ref{as}).
To better visualize this null cone, one could draw
(mentally or actually) in Fig.\ref{fig} 
the $\pm{\rm L_o}\sqrt{P}u$ asymptotes
from diagram (i) of  Fig.\ref{Fig}.
Integrating the electric flux ($\ast F$) through these $S^2[r]$ round spheres we find
\be
0=\int_{\cal V}\!d\ast F=\int_{\partial{\cal V}}\!\ast F=
\int_{\partial{\cal V_+}}\!\!\ast F+
\int_{\partial{\cal V_-}}\!\!\ast F=\left[4\pi Q\right]_{\cal S_+}
+\left[-4\pi Q\right]_{\cal S_-},
\label{div}
\ee
where the minus sign
comes from ${\partial{\cal V_-}}$ in the 
(moving backwards in time) ${\cal S_-}$, so
the overall null result is upheld. This leaves no
actual $Q$ to be trapped in any ${\cal V}$ in ${\cal S_-}\!\vee{\cal S_+}$;
the flux is not interrupted 
through any $S^2$ section, notably through
$S^2[r_o]$, so $E_C$  
is indeed a primordial field  in ${\cal G}$.
By the $\ast F\rightarrow F$ symmetry applied to (\ref{div}), $B_C=Q/r^2$ 
is likewise established as a primordial magnetic field, so there can be
no actual magnetic charge $Q$ or
Dirac-string singularities in any ${\cal G}$-geon monopole.
The smoothness of  potential and fields in (\ref{me})
cannot inhibit the emergence of  Dirac's quantization condition.
To see that explicitly, we turn to the $A^{\rm(m)}$ potentials
in (\ref{me}) for a pair of $A^{\rm(m)}_{\;\;\pm}$, to cover 
(as atlas with an equatorial overlap)
any given enclosing round-$S^2$, e.g.,
any typical $S^2[r]$ in Fig.\ref{fig}.
In our case, as with the Dirac-monopole,
exactly the same 
$2\pi n$ ($n\!\in\!Z$) phase difference will be recovered
in the mentioned $A^{\rm(m)}_{\;\;+}\cap A^{\rm(m)}_{\;\;-}$ overlap in ${\cal G}$.
And likewise for the ${\cal S}_Q$ geon, examined next.

\no
\section {Topological confinement of actual charges on $\partial{\cal S}_Q$}
\no
The concept of ${\cal S}_Q$ is referred-to as ${\cal G}/2$
because it emerged from
${\cal G}={\cal S_-}\!\vee{\cal S_+}$ 
when the $C^\infty$ junction
across $S^2[r_o]$ at $u=0$ was undone (severed),
leaving behind
a null squashed-$S^3$ boundary $\partial{\cal S}_Q\neq0$
as a physical singularity. 
As previewed, 
the new initial-value problem
involves  {\sc em}  field-lines terminating
on the round-$S^2[r_o]$ section of $\partial{\cal S}_Q$,
thus revealing actual electric (or magnetic) charge $Q$, 
trapped by topology and distributed
homogeneously on $S^2[r_o]$
as surface-charge density $\sigma_Q$.
To prove this, we can employ the divergence theorem with
$\sigma_Q$ as an ``almost point charge"
surrounded by any $S^2[r]$ as a Gauss sphere in Fig.\ref{fig}.
All incomplete geodesics {\em also} end 
on $\partial{\cal S}_Q$, so
the gravitational initial-value problem 
likewise uncovers the presence of
actual mass density $\rho_{m_S}$ on $S^2[r_o]$.
This will integrate to $m_S$
(cf.,  next section),  a mass-charge 
viewable as  bare mass  of ${\cal S}_Q$,
in full analogy to the $Q$-charge from  $\sigma_Q$.
The total mass of ${\cal S}_Q$ will be $2m_S$,
when we also include {\em effective} contributions from the
energy density of the surrounding gravitational and {\sc em} fields,
supplying an additional $m_S$ input.
Collecting these results, with $\rho_Q$ defined by analogy to 
$\rho_{m_S}$, we have
\be
\rho_Q=\sigma_Q\,\delta(r-r_o)=\frac{Q}{4\pi r^2}\,\delta(r-r_o),
\;\;\;\rho_{m_S}=\frac{m_S}{4\pi r^2}\,\delta(r-r_o),
\;\;\;\rho_{\rm em}=\frac{Q^2/r_o}{4\pi r^2}\,\delta(r-r_o),
\label{sigma}
\ee
where $Q^2/r_o$ is the {\sc em} potential self-energy
of $\sigma_Q$, confined as it is on $S^2[r_o]$. 
By the concept of any initial-value problem,
subsequent sections of ${\cal S}_Q$, propagating
beyond the `initial' ${\partial\cal S}_Q$,
will be totally `unaware' whether any detachment
has taken place at that ${\partial\cal S}_Q$.
Thus, an  overall stress-energy distribution $\tau_{\alpha\beta}$
must carry the full content of (\ref{sigma})
plus (not shown) contributions from higher moments, etc, in a
total energy-momentum distribution
\be
T_{\alpha\beta}= \tau_{\alpha\beta}\,\delta(r-r_o)\,.
\label{tau}
\ee
$T_{\alpha\beta}$, quite distinct from 
$T_{\alpha\beta}^{\rm ({\sc em})}$  in (\ref{ee}),
involves all mentioned or implied contributions from
charges  on $S^2[r_o]$,
namely electric (or magnetic) $Q$-charge and
higher moments, $m_S$ bare mass, $\kappa Q=2r_o$ diameter,
spin, gravitational
and {\sc em} self-energy, etc.
These are confined as `quantum numbers'
on the round $S^2[r_o]$,
which remains their host at  $r= r_o$
even if the initial-value problem is set 
on subsequent $S^2[r]$ sections
beyond ${\partial\cal S}_Q$: in spite
of accordingly large $r>r_o$ radii
in $(L_1)^2=(L_2)^2=r^2$
(cf., Fig.\ref{fig}), the $(L_3)^2=-(2r_o)^2$
value remains elementary, as a basic aspect of
topological confinement on $S^2[r_o]$
shared  by the general squashed-$S^3$  in  ${\cal S}_Q$.
This  could also relate to stability, as conjectured in the next section.

\no
\section {On asymptotic infinity, causality, and stability of
${\cal G}$}

All four ${\cal G}$,${\cal S}_Q$,$\bar{\cal G}$,$\bar{\cal S}_Q$
geons and
(defined by time-reversal)  antigeons
are asymptotically locally flat  manifolds
at $r\rightarrow  r_{\rm cl}$. 
The $r\rightarrow r_o$ limit is also
objective and physical, 
because it relates  to violations
of causality via
(\ref{bo}), as we'll see.
Vorticity is defined as
\be
\omega:=\ast(v\wedge dv)=\frac{2r_o}{r^2}\,\theta^3\,,
\label{o}
\ee
calculated here for an observer
with 4-velocity $V$, dual of
$v=dt+{\rm L_o}/{\sqrt{P}}\cos\theta d\phi$ from (\ref{m0}). 
As seen in Fig.\ref{Fig}, asymptotic infinity is actually
realized when $r$ has practically fallen on the 
$+{\rm L_o}\sqrt{P}|u|$ asymptote in (\ref{as}).
There, by (\ref{riemann}) etc, $R^\mu_{\;\nu\rho\sigma}$
vanishes at least as O$(r^{-2})$, with $W^\alpha_{\;\beta\gamma\delta}
\sim r^{-3}$. This  means that 
${\cal G}$,${\cal S}_Q$,$\bar{\cal G}$,$\bar{\cal S}_Q$
{\em could} carry spin ${\sf s}_G$, effective mass $m_G$,
and other charges, if found to be 
well-defined and finite  `quantum numbers', as mentioned.
They will then be shared
(up to $Q/|Q|$ signs)  by
${\cal G}$,${\cal S}_Q$,$\bar{\cal G}$,$\bar{\cal S}_Q$,
because all-four share the same asymptotic infinity.
To see that this is actually the case here,
and aiming to the upcoming (\ref{h}),
we introduce holonomic $x^i $ coordinates and
global $t$-time from (\ref{dt}) via (\ref{as}) as
\be
x^{\mu}=(x^0,x^i),\;\;\;\;\;
x^0=t={\rm L_o}{\sqrt{P}}u+2r_o\psi=
\pm\sqrt{r^2-r_o^2}+2r_o\psi\;\;\;\;
\left[\psi\,{\rm mod}\;4\pi n\right].
\label{t}
\ee
The $8\pi nr_o$ $(n\!\in\!Z)$ homotopy-group
structure from the $\psi\in[0,4\pi)$ angle
in the timelike dimension of $S^3$
is mandatory, so
any timelike direction in ${\cal G}$ 
can hardly avoid the involvement
of  the presence of $\psi$ and the causality-violating $t$-time loops it allows.
This potentially disastrous  result, which
also exists in Taub-NUT,
can here be naturally confined within sufficiently
{\em small} $r\geq r_o$ radii. These, even when
enormous w.r.t. Planck length,
can and must remain elementary.
Accordingly, classical causality is  protected
{\em if} the first bound in
\be
 r>r_{\rm sm}\geq r_o\;\;\;\;\;\;\left[|Q|\leq\frac{r_{\rm sm}}{\kappa}\right],
\;\;\;\;
\;\;\;\;\;m_G\geq\frac{|Q| }{\kappa}\;\longrightarrow\;\;\;P\leq8\pi^4
\label{bo}
\ee
can be observed [under a generally imposed constraint on the $Q$  parameter].
The second, a  Bogomol'nyi bound as it
applies in our case  \cite{gh},
has been evaluated  in terms of the effective $m_G$
from {\sc em} energy density via
the upcoming (\ref{mg});
it has been equivalently expressed as an upper
bound for the $P$ parameter. 
At $r\approx r_o$ scales,
this  Bogomol'nyi bound will be the only constraint applicable
so close to the Big Bang.
There\footnote 
{Where ${\cal G}_{\rm bb}$, as
a primordial pp-wave propagating out of
inflation, will be aiming toward classical scales.
},
the first bound in  (\ref{bo})
is violated for as long as the $2r_o\psi$ term 
in (\ref{t}) dominates over the
(normally enormous)  ${\rm L_o}\sqrt{P}u$ term,
as we'll see. 
Sufficiently beyond the $r\approx r_o$ region,
with $t$ turning null as $t\sim u$, 
classical causality is protected
and (\ref{rt}) holds 
for $t$ as well. At $r>> r_o$,
manifest general covariance in ${\cal G}$
can be traded for the standard 
$\eta_{\mu\nu}+h_{\mu\nu}$ perturbation in
Minkowski's  $M_{\!o}^4$, so,
at asymptotic infinity (if it acceptably exists,
with $h_{\mu\nu}\rightarrow 0$
as $r\rightarrow \infty$), $M_{\!o}^4$
is elevated to a $M^4\approx{\cal G}$.
Remarkably, here
we can actually have $M^4={\cal G}$  {\em exactly}.
Indeed, by (\ref{t}) etc, we can re-express (\ref{m0})
in such $\eta_{\mu\nu}+h_{\mu\nu}$ form in
terms of $(t,r,\theta,\phi)$ coordinates  as
\be
ds^2=\eta_{\mu\nu}dx^\mu dx^\nu
+\frac{r_o^2}{r^2-r_o^2}(dr)^2
-4r_o\cos\theta d\phi\left(dt+r_o\cos\theta d\phi\right),
\label{h}
\ee
to read-off  all $h_{\mu\nu}$, scaled as
they are by $\kappa Q=2r_o$. 
Thus , in addition to fixing the strength of the
{\sc em} content, the NUT-like charge 
also determines {\em where}
the gravitational asymptotic infinity
has actually been realized. 
Accordingly, the formal $r\rightarrow\infty$
limit can (and it will) be
safely replaced by an earlier one, e.g., the
$r\rightarrow   r_{\rm sm}$ in
the 4-scale hierarchy in (\ref{hi}).
The price for these deeper findings has been the
loss of manifest left-$SU(2)$ invariance, due to
the absorption of $\psi$ in the definition of $t$ back in (\ref{dt}),
here realized  as the survival of $\theta,\phi$ in (\ref{h}).
This could (and here it does) hinder the calculation of
mass and spin \cite{w} (p.165~ff). Accordingly, we have to resort to 
estimates. Thus, to evaluate the
Bogomol'nyi bound in (\ref{bo}), we assume
that $m_G=2m_S$ comes solely from {\sc em} energy density.
Integrating $T_{00}^{\rm (em)}$
between $r_o$ ($u=0$) and 
$r$, with volume element 
$\theta^1\wedge\theta^2\wedge\theta^3$,
we find a finite $m_S$ value
\be
m(r)=\pi Q^2\sqrt{\frac{2}{P}}\int_{r_o}^r\frac{d\sqrt{r^2-r_o^2}}{r^2}
\;+O(r^{-2})\,\;\;\stackrel{r\rightarrow\infty}{\longrightarrow}\;\;\;\;
m_S=\frac{\pi^2 Q^2}{\sqrt{2P}r_o},
\label{mi}
\ee
at the formal $r\rightarrow\infty$ limit.
Practically, this $m_S$ value
(and asymptotic infinity as $h_{\mu\nu}\rightarrow 0$) has been
already reached at much-earlier limits,
e.g., the $r\rightarrow  r_{\rm sm}$
in the case of a 4-scale hierarchy in (\ref{hi}).
We also note that, had the geometry
allowed the $r_o=0$ value,
the $r\rightarrow\infty$  limit in (\ref{mi}) 
would have simply reproduced the notorious
(and here disastrous) result of a diverging $m_S$.
By integrating over $u\in(-\infty,+\infty)$  in (\ref{mi}) to
cover the entire ${\cal G}$ manifold
(and likewise with $\omega r^2dm$ as angular momentum
element), we find 
\be
r\longrightarrow  r_{\rm sm}\;\;\left(h_{\mu\nu}\longrightarrow 0\right):\;\;\;\;\;\;
\;\;\;\;m_G=2m_S\approx\frac{2\sqrt{2}\pi^2 Q^2}{\rm L_o}\,,\;\;\;\;\;
{\sf s}_G\sim\frac{2\sqrt{2}\pi^2 Q^2}{\sqrt{P}}\,,
\label{mg}
\ee
etc, where (\ref{o}) has also been used to estimate spin. 
We recall that these results are 
shared as `quantum numbers' by all four
${\cal G}$,$\bar{\cal G}$,${\cal S}_Q$,$\bar{\cal S}_Q$,
up to a $Q/|Q|$ sign and particular aspects.
An example of the latter is the $2m_S$ value
as mass of ${\cal S}_Q$, realized as the sum
of $m_S$ as the bare mass in (\ref{sigma}), 
plus the {\sc em} contribution  (an additional $m_S$)  from 
integrating over a single covering of ${\cal S}_Q$,
actually as calculated in (\ref{mi}).

There exists no interaction
between the ${\cal S_\pm}$ constituents
of ${\cal G}={\cal S_-}\vee{\cal S_+}$, hence
neither a relation to $\bar pp$ (positronium-like)
states, which are typically unstable.  
A plausible and in agreement with 
observation (but only comparative) statement on the stability of
the ${\cal G}$ geon is that the magnetic 
(\ref{me}) types are favored
vs the electric (\ref{em}), as the former
have very few or virtually no channels to
decompose into conventional
magnetic monopoles or disperse into magnetic vortices.
The confinement of actual charge on $S^2[r_o]$
could relate to the stability of ${\cal S}_Q$,
if the latter could be viewed as an equilibrium
state  between  gravitational collapse within $S^2[r_o]$
and the outburst of $\sigma_Q$ off $S^2[r_o]$
as  no-go extremes.
Similar approaches do exist, but they are
all tentative prior to a needed rigorous study.
In any case, any issue or result on the stability of ${\cal G}$ 
would also illuminate 
the likewise suspended issue on the stability
of the Kerr-Newman solution, which has been fundamentally
upgraded by ${\cal S}_Q$ .


\section {Conclusions}

1. We have examined the strong similarities and
the even  stronger differences
between the ${\cal G}$-geon vs the Taub-NUT.   
Sections 4,5 also allow a
comparison between the ${\cal S}_Q$-geon
and the Kerr-Newman solution,
with the  content of the former being richer
(with higher {\sc em} moments,
in addition to spin, mass, etc)
and more predictive\footnote
{These findings cannot apply to Reissner-Nordstr\"om etc,
because the spin of ${\cal G}$ or
${\cal S}_Q$  cannot vanish.}. 
${\cal S}_Q$ is also non-singular,  regardless of hierarchy type in (\ref{hi}),
as any geon must be by concept.
The admittance of any spacetime singularity
sufficiently close to the Big-Bang
would redirect the latter's dynamics (one of a time-reversed black hole) 
toward that of the added singularity, so as to
produce a loop of failed or aborted Big-Bang,
roughly as a wormhole with
its open regions topologically identified.
${\cal G}$,${\cal S}_Q$,$\bar{\cal G}$,$\bar{\cal S}_Q$,
referred-to  collectively
as ${\cal G}$ unless explicitly distinguished, 
share the same asymptotic infinity, hence
the same spin, mass, etc, up to $Q/|Q|$ signs.
They also admit the  presence of a
sufficiently {\em weak} scalar field 
in a 3-parameter generalization\footnote
{Or 4-parameter Einstein-Maxwell-Yang-Mills exact solutions
with no external sources (work in progress).}.

\no
2. ${\cal G}$ geons admit timelike loops,
bifurcating geodesics, etc, but,
as a fifth fundamental difference vs Taub-NUT,  
the natural confinement of such non-classical dynamics\footnote{ 
Dynamics which clearly hints at  a
quantum-gravity environment, and should be accordingly delimited.
}
has been possible within the bounds
described by the first inequality in (\ref{bo}). Whenever the latter
can be observed, accordingly-enforced 
is the protection of classical causality. The
violation of the first constraint in (\ref{bo}), 
hence of causality, is of course an important aspect
of the quantum regime. Here however, this violation acquires
additional importance for models
close to the Big-Bang, as with the
${\cal G}_{\rm bb}$ geon (cf., paragraph 5 below). 

\no
3. The constraints in (\ref{bo}) are well-defined in ${\cal G}$,
but, particularly  the  Bogomol'nyi bound,
cannot be really applied to the Taub-NUT vacuum.
Expressed as
upper-bounds on $P,Q$  in (\ref{bo}), they also shape
as accordingly-constrained the $(P,Q)$ parameter space of ${\cal G}$.
With no constraints, this space would involve
any $P>0$, $Q\neq0$ value, with 
(i) $Q$ fixing $2r_o=\kappa |Q|$,
hence the asymptotic infinity for gravity
and the strength of the {\sc em} content 
(effective or actual), and
(ii) $P/Q^2$  providing independent
scaling of the $m_G$ mass in units of $\kappa$. 
The constraints in (\ref{bo}) are also
expected to incite predictions, as it actually
happens in the following previews on
anticipated ${\cal G}$-geon dynamics.

\no
4.  The main idea and approach 
is to describe and study this dynamics in terms of 
analytic simulations (discreetly distanced from controversial
HEP considerations), in Friedman-like evolution models
${\cal F}$, one per case, such as the ${\cal F}_o$ and ${\cal F}'_o$
(outlined next) or the ${\cal F}_{\rm bb}$
(outlined last).
To begin with the simplest non-trivial example, 
(i) ${\cal F}_o$  is filled with
${\cal G}_o$ geons, hardly interacting to simulate dark-matter dust. 
The $ r_{\rm cl}$ radius in (\ref{hi})
can be exploited here as a third parameter, so the
$P,Q$ can be spent to restrain mass and $Q$-charge
of the DM particle within bounds compatible with 
data from HEP and cosmology. Then,
spin, dipole and quadrapole moments, as well as the
dating of ${\cal G}_o$ in
early afterglow  in a
4-scale  hierarchy (\ref{hi}), 
would then be predictions of the model.
(ii) ${\cal F}'_o$ is also filled
with ${\cal G}_o$ geons, now 
mixed randomly with ${\cal G}'_o$ geons which are fewer but
have much larger $Q'$,$m'_G$ values compared to
those of ${\cal G}_o$. Having 
$P'$,$Q'$ values in a 3-scale hierarchy in (\ref{hi}), the ${\cal G}'_o$
as seed particles will  expectedly shape the dynamics in ${\cal F}'_o$ as one
of accretion, trapping DM particles plus
baryons (if also present in the model).
A primordial stability-enhancing magnetic field is
actually predicted as $B'_C=Q'/{r'_{\rm cl}}^{\!2}$ at the time
this pre-galactic dynamics commences (near recombination,
around the end of the so-called dark ages).

\no
5. The ${\cal F}_{\rm bb}$ model could involve many or {\em just one} 
`cosmogonic' ${\cal G}_{\rm bb}$ geon in a
4-scale hierarchy in (\ref{hi}), and
an (even sub-Planck) $2r_o$ size.
${\cal G}_{\rm bb}$ could provide analytic simulations of 
primordial gravitational waves, created with
the first quantum fluctuation(s) of the vacuum.
These
configurations are highly non-linear and
must be in agreement with the  Bogomol'nyi  bound in (\ref{bo}),
hence with possibly enormous
`mass-energy'. At the same time, 
the first bound in (\ref{bo}) {\em must} be violated
for as long as $t$  in (\ref{t}) is sufficiently close to $t=0$,
or, equivalently, $r$ is extremely near $r_o$. There, large
$n>1$ values are allowed and they could even be induced in
repeatedly circling time-loops,
feeding superluminal  expansion and
{\em global} violation of  causality, in
exact models for analytic simulations
of  inflationary  dynamics. This dynamics, 
with amplifications etc, is expected to last
for as long as the $2r_o\psi$ term retains its dominance
over the $\sqrt{r^2-r_o^2}$ term in (\ref{t}).
When that dominance is reversed by sufficiently large $r$,
the first bound in (\ref{bo}) is realized as applicable
for the first time, inflation stops, and an
expanding {\em causal} classical regime is born.
The ${\cal G}_{\rm bb}$ geon(s), created 
shortly after the Big Bang and
amplified during inflation, are expected to
reach asymptotic infinity somewhere
within the afterglow era. As a general result, the
${\cal G}_o$,${\cal G}'_o$,${\cal G}_{\rm bb}$ 
examples also serve as paradigms of
gravity giving $2r_o$  size from NUT-like charge,
and independently-scaled  mass (or pp-wave energy) to
particle-like configurations. Those previewed here, as well as
other ${\cal G}$ or ${\cal S}_Q$ geon configurations,
may provide analytic models with current observational
and novel theoretical interest 
in particle physics and cosmology.

The author is grateful to A.A. Kehagias for discussions.


\vskip .4in
\begin{thebibliography}{99}

\bibitem{e}
A.~Einstein, L. Infeld, B.~Hoffmann, {\it Ann. of Math. (2nd series)} {\bf 39} (1938) 65-100;
A.~Einstein and W.~Pauli, Ann. Math. {\bf 44} (1943) 131;
A.~Papapetrou, {\it Ann. Physik.} {\bf 9} (1962) 97.
\bibitem{c}
\'E.~Cartan, {\it La th\'eorie des groupes finis et continus et la g\'eom\'etrie
diff\'erentielle trait\'ees
par la m\'ethode du rep\'ere mobile} (Gauthier-Villars, Paris 1937).
\bibitem{l}
A.~Lichnerowicz, {\em Th\'eories Relativiste de la Gravitation et de l'\'Electromagn\'etisme}
\\(Masson, Paris, 1955).
\bibitem{wh}
J.A.~Wheeler, {\it Phys.Rev.} {\bf 97} (1957) 511.
\bibitem{t}
A.H.~Taub, {\it An. Math.} {\bf 53} (1951) 472;
E.~Newman, L.~Tamburino, T.~Unti, {\it J. Math. Phys.} {\bf 4} (1963) 915;
C.W.~Misner, {\it J. Math. Phys.} {\bf 4} (1963) 924;
C.W.~Misner and A.H.~Taub (1968), {\it Engl. Transl. Sov. Phys.-JETP}
{\bf 28} (1969) 122.
\bibitem{rs}
M.~Ryan and L.~Shepley, {\it Homogeneous Relativistic Cosmologies} 
(PUP, N.J., 1975); T.~Eguchi. P.B.~Gilkey, A.J.~Hanson,
{\it Phys.Rep.} {\bf 66} (1980) 213-393;
M.J.~Duff, B.E.W.~Nilsson, C.N.~Pope, {\it Phys. Rep.} {\bf 130} (1986) 1.
\bibitem{gp}
J.P.~Griffiths and J.~Podolsky, {\it Exact space-times in Einstein's General Relativity}\\ 
(Cambridge Monographs on Mathematical Physics, Cambridge 2010).
\bibitem{ab}
D.R.~Brill and J.B.~Hartle, {\it Phys.Rev.}~{\bf 135} (1964) B271;
P.R.~Anderson and D.R.~Brill {\it Phys.Rev.D} {\bf 56} (1997) 4824.
\bibitem{so}
R.D.~Sorkin, {\it Introduction to topological geons} 
(Erice, Italy, 1985); J.~Louko,
{\it J. Phys.}: Conf. Ser. 222 012038 (2010).
\bibitem{gh}
G.W.~Gibbons and C.M.~Hull,
{\it Phys. Lett. B} {\bf 199} (1982) 190.
\bibitem{w}
S.~Weinberg, {\it Gravitation and Cosmology} 
(Wiley, New York, 1972).

\end{thebibliography}
\end{document}